\documentclass[draft,nofootinbib,aps,pre,twocolumn,showpacs,groupaddress,preprintnumbers,floatfix]{revtex4-1}

\usepackage{dynlearn}
\usepackage{color}
\usepackage{amsmath}
\usepackage{cleveref}

\begin{document}

\def\ourTitle{%
Extreme Quantum Advantage when Simulating Strongly Coupled Classical Systems
}

\def\ourAbstract{%
Classical stochastic processes can be generated by quantum simulators instead
of the more standard classical ones, such as hidden Markov models. One reason
for using quantum simulators is that they generally require less memory than
their classical counterparts. Here, we examine this quantum advantage for
strongly coupled spin systems---the Dyson-like one-dimensional Ising spin
chain with variable interaction length. We find that the advantage scales with
both interaction range and temperature, growing without bound as
interaction increases. Thus, quantum systems can very efficiently simulate
strongly coupled classical systems. 
}

\def\ourKeywords{%
stochastic process, Ising model, Dyson model, long-range interaction, strongly
correlated system, quantum information, quantum communication, quantum
complexity, quantum machine, statistical complexity
}

\hypersetup{
  pdfauthor={Cina Aghamohammadi},
  pdftitle={\ourTitle},
  pdfsubject={\ourAbstract},
  pdfkeywords={\ourKeywords},
  pdfproducer={},
  pdfcreator={}
}

\author{Cina Aghamohammadi}
\email{caghamohammadi@ucdavis.edu}
 
\author{John R. Mahoney}
\email{jrmahoney@ucdavis.edu}

\author{James P. Crutchfield}
\email{chaos@ucdavis.edu}

\affiliation{Complexity Sciences Center and Physics Department,
University of California at Davis, One Shields Avenue, Davis, CA 95616}

\date{\today}
\bibliographystyle{unsrt}

\title{\ourTitle}

\begin{abstract}
\ourAbstract
\end{abstract}



\preprint{\sfiwp{16-09-XXX}}
\preprint{\arxiv{1609.XXXX}}

\title{\ourTitle}
\date{\today}
\maketitle

\setstretch{1.1}

\newcommand{\SpinProcess}{\text{P}}
\newcommand{\kB}{k_B}

\section{Introduction}

We illustrate, what seems to be, an emerging principle relating the classical
and quantum worlds: strongly correlated classical systems can be efficiently
simulated by quantum systems. Moreover, we show that this quantum advantage is
substantial and increases with the classical system's degree of interaction.

Our results suggest that this principle may well lead to broad consequences.
The world is nothing, if not its constituent interactions. Some go so far as to
argue that \emph{interaction}, not object, is the basic unit of reality
\cite{Whit78a,Ross06a}. At a minimum, though, interactions make our world
interesting and structured. They also complicate it, and rapidly as the web of
interaction widens. The resulting complicatedness often confounds theory and
drives us to resort to simulation.

Statistical mechanical spin systems are a prime example: A broadly used class
of models that can be arbitrarily hard to analyze and, somewhat soberingly,
even hard to simulate \cite{Ande89a,Stei13a}. While there is a plurality of
complicated spin systems, we choose one that is sufficiently interesting, but
also very familiar territory: the one-dimensional Ising spin chain with
variable interaction range. The one-dimensionality provides tractability, while
the long-range coupling yields the desired complicated configuration structure.

Selecting a system is only the start, of course. There are many different
simulation methods for any given class. The differences range from the purely
algorithmic to the physical substrate employed. For instance, a computer can
use machine code or high-level programming languages to yield the desired
result. Additionally, computers come in different varieties, in particular the
substrate may be classical or quantum \cite{Feyn82a}. Just as the choice of
code influences the computational resources required, so does the substrate.

Here, we report on the memory resources required by classical and quantum
simulators. It is known that a quantum simulator typically requires less memory
than its classical counterpart.\footnote{The memory resources required are
equal if and only if the classical stochastic process is that with no
crypticity \cite{Maho16}.} We refer to this as the \emph{quantum advantage}.
Despite exploring several particular cases, very little is known about how this
quantum advantage scales. Addressing this, the following shows that not only is
the advantage substantial, it also increases systematically and without bound.

To establish the scalings we compare classical and quantum simulators that
generate spin configurations of the one-dimensional Ising model with
$N$-nearest neighbor interactions. To compute the quantum advantage, we adapt
the transfer-matrix formalism to construct the two simulators. This technique
allows us to numerically (but accurately) determine the scaling behavior. We
find that not only is the quantum advantage generic, its growth 
scales with temperature and interaction length.

\section{Dyson-Ising Spin Chain}

We begin with a general one-dimensional ferromagnetic Ising spin chain
\cite{BAXTER07,Agha12} with Hamiltonian:
\begin{align*}
\mathcal{H}= -\sum_{\langle i,j \rangle} J(i,j) s_is_j~,
\end{align*}
in contact with thermal bath at temperature $T$,\footnote{Throughout, $T$
denotes the effective temperature $\kB T$.} where spin $s_i$ at site $i$
takes on values $\{ +1,-1\}$ and $J(i,j) \geq 0$ is the spin coupling constant
between sites $i$ and $j$. Assuming translational symmetry, $J(i,j) \to J(k),~k
\equiv |i - j|$. Commonly, $J(k)$ is a positive and monotone-decreasing
function. An interaction is said to be \emph{long-range} if $J(k)$ decays more
slowly than exponential. In our studies, we consider couplings that decay by a
power law:
\begin{align*}
J(k)= \frac{J_0}{k^\delta}
  ~,
\end{align*}
where $\delta>0$. The spin chain resulting from these assumptions is called the
\emph{Dyson} model \cite{Dyson69}.

To approximate such a long-range system one can consider a similar system with
finite-range interaction. For every interaction range $N$, we
define the approximate Hamiltonian:
\begin{align*}
\mathcal{H}_N= -\sum_{i} \sum_{k=1}^{N}\frac{J_0}{{k}^\delta} s_is_{i+k}
  ~.
\end{align*}
This class of Hamiltonians can certainly be studied in its own right, not
simply as an approximation. Let's explore its set of equilibrium configurations
as a stochastic process.

\newcommand{\cs}{\causalstate}
\newcommand{\CS}{\CausalState}
\renewcommand{\H}{\operatorname{H}}

\section{Processes}

The concept of a stochastic process is very general. Any physical system that
exhibits stochastic dynamics in time may be thought of as \emph{generating} a
stochastic process. We focus on \emph{discrete-time, discrete-valued stationary
stochastic processes}. Such a process, denoted $\SpinProcess =
\big\{\MeasAlphabet^\infty, \Sigma, \mathbb{P}(.) \big\}$, is a probability
space \cite{Uppe97a,Travers13}. Here, the observed symbols come from an
alphabet $\MeasAlphabet = \{\downarrow,\uparrow\}$ of local spin states; though
our results easily extend to any finite alphabet. Each random spin variable
$\MeasSymbol_i, ~i \in \mathbb{Z},$ takes values in $\MeasAlphabet$.
$\mathbb{P}(.)$ is the probability measure over the bi-infinite chain of random
variables $\MeasSymbol_{-\infty:\infty} = \ldots \MeasSymbol_{-2}
\MeasSymbol_{-1} \MeasSymbol_0 \MeasSymbol_1 \MeasSymbol_2 \ldots$. $\Sigma$ is
the $\sigma$-algebra generated by the cylinder sets in $\MeasAlphabet^\infty$.
\emph{Stationarity} means that $\mathbb{P}(.)$ is invariant under time
translation. That is, $\mathbb{P}(\MeasSymbol_{i_1}\MeasSymbol_{i_2} \cdots
\MeasSymbol_{i_m}) = \mathbb{P}(\MeasSymbol_{i_1+n}\MeasSymbol_{i_2+n} \cdots
\MeasSymbol_{i_m+n})$, for all $m \in \mathbb{Z}^+$ and $n \in \mathbb{Z}$.

To interpret our Ising system as a stochastic process, we consider not
its time evolution, but rather the spatial ``dynamic''. A spin configuration at
one instant of time may be thought of as having been generated left-to-right
(or equivalently right-to-left). The probability distribution over these
configurations defines a stochastic process $\SpinProcess (N,T)$ that
inherits its stationarity from spin-configuration spatial translation
invariance. In this way, we build on earlier work that used computational
mechanics to analyze statistical structure in spatial configurations generated
by spin systems \cite{Crut97a,Feld98b}.

Now that we have defined the process of interest, let us introduce two of its simulators.

\section{Classical and Quantum Simulators}

\begin{figure}
\includegraphics[width=1\linewidth]{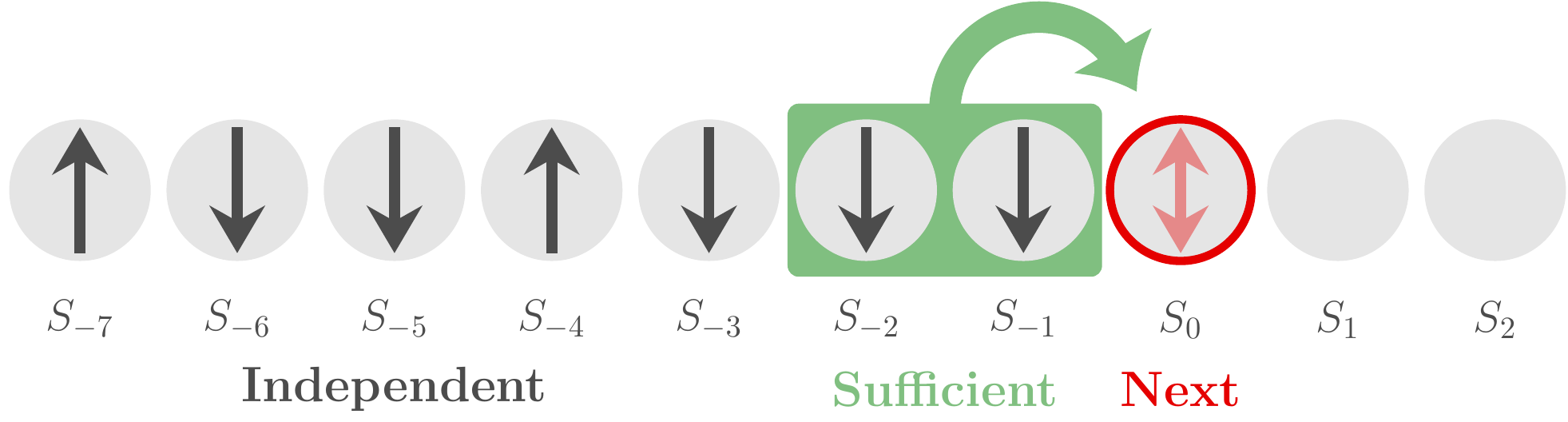}
\caption{A Markov order-$N$ process generates a spin configuration from
	left-to-right. Markov order $N = 2$ shown. The values of an isolated spin
	$S_0$, say, is undetermined. To make this (stochastic) choice consistent
	with the overall process and the particular instantiation on the left, it
	is sufficient to consider only the previous $N$ ($2$) spins (highlighted in
	green).
  }
\label{fig:Markov_R_spin}
\end{figure}

What is a simulator for a stochastic process? Often, ``simulation'' refers to
an approximation. In contrast, we require our simulators to be perfect, to
generate $\SpinProcess$'s configurations and their probabilities exactly. Our
simulator, though, does more than correctly reproduce a probability
distribution over bi-infinite configurations. Specifically, a \emph{simulator}
$S$ of process $\SpinProcess$ is an object where, given an instance of a
semi-infinite ``past'' $\past = \ldots, \meassymbol_{-3}, \meassymbol_{-2},
\meassymbol_{-1}$, a query of $S$ yields a sample of the ``future'' $\Future =
\meassymbol_0, \meassymbol_1, \ldots$ from the conditional probability
distribution $\Prob(\Future | \Past = \past)$. See Fig. \ref{fig:Markov_R_spin}.

Physical systems, under certain assumptions such as thermal equilibrium, manifest
stationary stochastic processes.
When we refer to the simulation of a physical system, what we mean is the 
simulation of these processes.

How are these simulators implemented? Two common formalisms are \emph{Markov
Chains} \cite{Norr98,LEVIN09} and \emph{Hidden Markov Models} (HMM)
\cite{Rabi86a,Rabi89a,Uppe97a}. The latter can be significantly more compact in
their representation\footnote{For example, the Simple Nonunifilar Source
(SNS) process requires an infinite-state Markov chain and consequently requires
infinite memory for simulation, while its HMM requires only two states and so
a single bit of memory \cite{Marz14b}.} and, for this reason, are sometimes the
preferred implementation choice. Here, we employ a particularly useful form of
HMM generators.

These HMMs represent the generating mechanism for a given process by a tuple
$\big\{ \CausalStateSet,  \MeasAlphabet, \{ T^{(\meassymbol)}: \meassymbol \in
\MeasAlphabet \}, \big\}$ where $\CausalStateSet$ is a finite set of states
called \emph{causal states}, $\MeasAlphabet$ is a finite discrete alphabet and
$\{T^{(\meassymbol)}: \meassymbol \in \MeasAlphabet\}$ are $|\CausalStateSet|
\times |\CausalStateSet|$ substochastic symbol-labeled transition matrices. The
latter's sum $\mathbf{T}= \sum_{\meassymbol \in \MeasAlphabet}
T^{(\meassymbol)}$ is a stochastic matrix. A \emph{unifilar} HMM is one in
which each row of each substochastic matrix has at most one nonzero
element.\footnote{A fledgling literature on minimal \emph{nonunfilar} HMMs
\cite{Lohr09b} exists, but constructive methods are largely lacking and, as a
consequence, much less is known \cite{Lohr09c,Lohr12,Gmei11a}.}

\paragraph*{\textbf{\EM}}
A given stochastic process can be correctly generated by any number of unifilar
HMMs. The one requiring the minimum amount of memory for implementation is
called the \emph{\eM} \cite{Crut12a} and was first introduced in Ref.~\cite{Crut88a}. 
A process' \emph{statistical complexity} $\Cmu$ \cite{Crut12a}
is the the Shannon entropy of the \eM's stationary state distribution: $\Cmu =
\H(\CausalState) = - \sum_{\cs \in \CausalState} \Pr(\cs) \log_2 \Pr(\cs)$. Key
to our analysis of classical simulator resources, it measures the minimal
memory for a unifilar simulator of a process. $\Cmu$ has been determined for a
wide range of physical systems
\cite{Perr99a,Delg97a,Neru10,Nerukh12,Kell12a,Li13a,Varn14a}. Helpfully, it and
companion measures are directly calculable from the
\eM, many in closed-form \cite{Crut13a}.

\paragraph*{\textbf{Ising \eM}}
How do we construct the \eM\ that simulates the process $\SpinProcess(N,T)$?

First, we must define process' \emph{Markov order} \cite{LEVIN09}: the minimum
history length $R$ required by any simulator to correctly continue a
configuration.\footnote{More precisely, we mean that an ensemble of simulators
must be able to yield an ensemble of configurations that agree (conditioned on
that past) with the process' configuration distribution.} Specifically, $R$ is
the smallest integer such that:
\begin{align*}
\mathbb{P}(\MeasSymbol_t | \ldots, \MeasSymbol_{t-2}, \MeasSymbol_{t-1}) = \mathbb{P}(\MeasSymbol_t | \MeasSymbol_{t-R}, \ldots, \MeasSymbol_{t-2}, \MeasSymbol_{t-1})~.
\end{align*}

Reference \cite[Eqs. $(84)-(91)$]{Feld98b} shows that $\SpinProcess(N,T)$ has
Markov order $N$ for any finite and nonzero temperature $T$. One concludes that
sufficient information for continued generation is contained in the
configuration of the $N$ previously generated spins. More importantly, the \eM\
that simulates $\SpinProcess (N,T)$ has $2^N$ states and those states are in
one-to-one correspondence with the set of length-$N$ spin configurations.

Second, another key process characteristic is its \emph{cryptic order}
\cite{Maho09a,Maho11a}: the smallest integer $K$ such that $\H(\CausalState_K |
\MeasSymbol_{0}, \MeasSymbol_{1} \ldots)=0$, where $\H[W|Z]$ is the conditional
entropy \cite{Cove06a} and $\CausalState_K$ is the random variable for the
$K^\text{th}$ state of the \eM\ after it generated symbols $\MeasSymbol_{0},
\MeasSymbol_{1} \ldots$. Using the fact that \eM\ states are in one-to-one
correspondence with the set of length-$N$ spin configurations, it is easy to
see that $\SpinProcess(N,T)$'s cryptic order $K = N$.

\begin{figure}
\centering
\includegraphics[width=\linewidth]{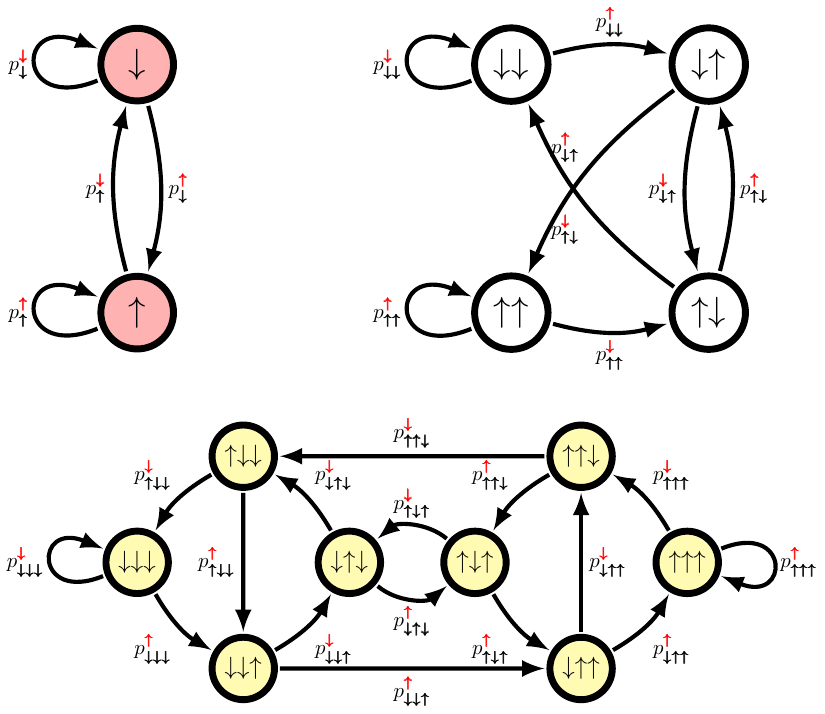}
\caption{\EM\ generators of 1D-configuration stochastic processes in
	Dyson-Ising systems of increasing correlational complexity ($N = 1, 2,
	3$): $\SpinProcess(1,T)$ (top left), $\SpinProcess(2,T)$ (top right) and
	$\SpinProcess(3,T)$ (bottom).
  }
\label{MARKOVIN}
\end{figure}

Figure~\ref{MARKOVIN} shows the unifilar HMM generators (\eMs) of the processes
$\SpinProcess(N,T)$ for $N=1, 2,$ and $3$. Let's explain.  

Consider the spin process $\SpinProcess(1,T)$, a Markov-order $R = 1$
process. To generate the process we only need to remember the last spin
generated. The \eM\ (Fig.~\ref{MARKOVIN} top-left) has two states, $\uparrow$
and $\downarrow$. If the last observed spin is $\uparrow$, then the current
state is $\uparrow$ and if it is $\downarrow$, the current state is
$\downarrow$. We denote the probability of a $\downarrow$ spin given a previous
$\uparrow$ spin by $p^{{\color{red}\pmb\downarrow}}_{\pmb{\uparrow}}$. The
probability of an $\uparrow$ spin following a $\uparrow$ spin is the complement.

Consider the process $\SpinProcess(2,T)$ with Markov-order $R = 2$
and so longer-range interactions. Sufficient information for generation is
contained in the configuration of the two previously generated spins. Thus, the
\eM\ (Fig.~\ref{MARKOVIN} top-right) has four states that we naturally label
$\uparrow\uparrow$, $\uparrow\downarrow$, $\downarrow\uparrow$, and
$\downarrow\downarrow$. If the last observed spin pair $x_{-1}x_0$ is
$\uparrow\downarrow$, the current state is $\uparrow\downarrow$. Given this
state, the next spin will be $\uparrow$ with probability
$p^{{\color{red}\pmb\uparrow}}_{\pmb{\uparrow\downarrow}}$ and $\downarrow$
with probability $p^{{\color{red}\pmb\downarrow}}_{\pmb{\uparrow\downarrow}}$.
Note that this scheme implies that each state has exactly two outgoing
transitions. That is, not all transitions are allowed in the \eM.

Having identified the state space, we may calculate the \eM\ transition
probabilities $\{ T^{(\meassymbol)} \}_{ \meassymbol \in \MeasAlphabet}$. We
first compute the transfer matrix $\mathbf{T}$ \cite{Dobs69} and then extract
conditional probabilities, following Ref. \cite{Feld98b}. (See
App.~\ref{sec:AnalyticalResult} for details.)
The statistical complexity $\Cmu$ follows straightforwardly from the \eM.

\paragraph*{\textbf{q-Machine}}
By studying a specific process (similar to the \eM\ in top-left of
Fig.~\ref{MARKOVIN}), Ref.~\cite{Gu12a} recently demonstrated that quantum
mechanics can simulate stochastic processes using memory capacity less than
$\Cmu$. This motivates a search for more efficient quantum simulators of other
stochastic processes.

A class of such simulators, called \emph{q-machines}, applicable to arbitrary
processes, was introduced in Ref.~\cite{Maho16}. This construction depends on
an encoding length $L$, each with its own q-machine and its quantum cost
$C_q(L)$. The cost $C_q(L)$ saturates at a particular length, which was shown
to be the process' cryptic order $K$, introduced above \cite{Maho11a}. And so,
we restrict ourselves to this choice ($L = K$) of encoding length and refer
simply to the q-machine and its cost $C_q$.

The q-machine's quantum memory $C_q$ is upper-bounded by $\Cmu$, with equality
only for the special class of zero-cryptic-order processes \cite{Maho11a}.  And
so, $\Cmu / C_q$ gives us our quantitative measure of \emph{quantum advantage}.
Efficient methods for calculating $C_q$ were introduced by Ref.~\cite{Riech16a}
using spectral decomposition. Those results strongly suggest that the q-machine
is the most memory-efficient among all unifilar quantum simulators, but as yet
there is no proof.\footnote{As in the classical case, nonunifilar quantum
simulators are much less well understood \cite{Monr16, Gmei11a, Monr10a}.}
The quantum advantage $\Cmu / C_q$ has been investigated both analytically
\cite{Maho16,Riech16a,Agha16a,Suen15a,Tan14} and experimentally \cite{Pals15}.

\begin{figure}
\includegraphics[width=\linewidth]{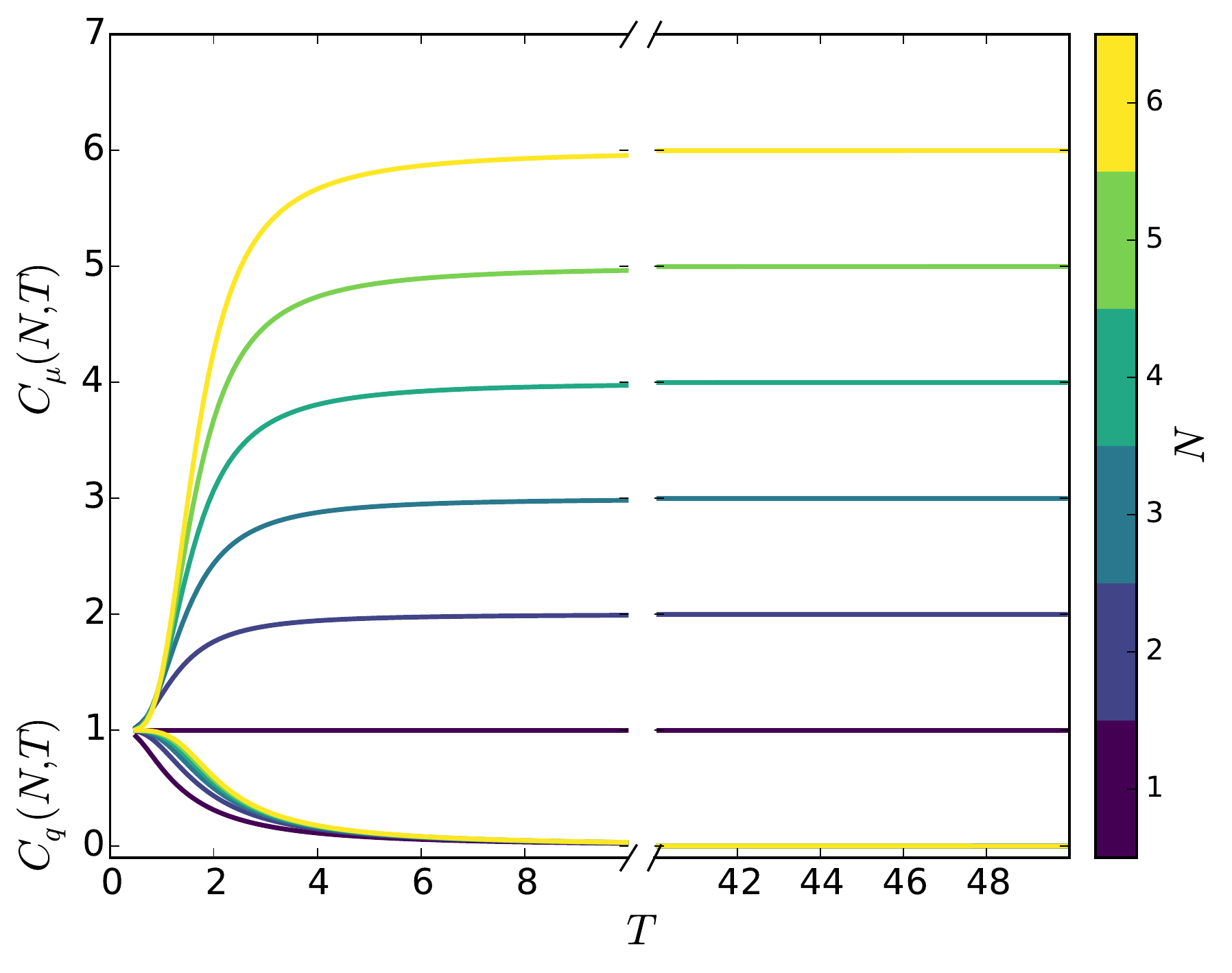}
\caption{Classical memory $\Cmu(N,T)$ and quantum memory $C_q(N,T)$
	required for simulating the Dyson-Ising processes $\SpinProcess(N,T)$ for
	$N=1, \ldots, 6$, a range of temperatures $T = 1, \ldots, 50$ and
	$\delta=2$. Note the dramatic difference in behaviors. $\Cmu(\cdot)$ and
	$C_q(\cdot)$ both are increasing functions of $N$. However, $\Cmu(\cdot)$
	is an increasing function of $T$ and $C_q(\cdot)$ is decreasing function of
	$T$ and bounded by $1$, independent of $N$ and $T$.
	}
\label{fig:Cmu_Cq_v_T_for_many_N}
\end{figure}

The q-machine is straightforward to construct from a given \eM. It consists of
a set $\{ \ket{\eta_i} \}$ of pure quantum \emph{signal states} in one-to-one
correspondence with the classical causal states $\causalstate_i \in
\CausalStateSet$. Each signal state $\ket{\eta_i}$ encodes the set of
length-$K$ (cryptic order) sequences that may follow $\causalstate_i$, as well
as each corresponding conditional probability: 
\begin{align*}
\ket{\eta_i} \equiv
  \sum \limits_{w \in |\MeasAlphabet|^K}
  \sum \limits_{\causalstate_j \in \CausalStateSet}
  {\sqrt{\mathbb{P}(w, \causalstate_j | \causalstate_i)}
  ~ \ket{w} \ket{\causalstate_j}}
  ~,
\end{align*}
where $w$ denotes a length-$K$ sequence and $\mathbb{P}(w,\causalstate_j |
\causalstate_i) = \mathbb{P}(X_{0} \cdots X_{K-1} = w,\CausalState_{K-1} =
\causalstate_j | \CausalState_0 = \causalstate_i)$. The resulting Hilbert space
is the product $\mathcal{H}_w \otimes \mathcal{H}_{\sigma}$. Factor space
$\mathcal{H}_{\sigma}$ is of size $|\mathcal{S}|$, the number of classical
causal states, with basis elements $\ket{\causalstate_i}$. Factor space
$\mathcal{H}_{w}$ is of size $|\mathcal{A}|^K$, the number of length-$K$
sequences, with basis elements $\ket{w} = \ket{x_0} \cdots \ket{x_{K-1}}$.
For $\SpinProcess(N,T)$'s \eMs, $|\mathcal{S}|=2^N$ and $|\mathcal{A}|^K=2^N$.
The q-machine's density matrix $\rho$ is defined by:
\begin{align}
\label{rho}
\rho= \sum \limits_i {\pi_i \ket{\eta_i} \bra{\eta_i}}
 ~,
\end{align}
where $\{\pi_i\}$ is the stationary distribution over the \eM's states.
This is determined from the left eigenvector of transfer matrix $\mathbf{T}$
corresponding to the eigenvalue $1$. From the density matrix $\rho$, it is
straightforward to calculate $C_q = S(\rho)$---$\rho$'s von Neumann entropy.

\section{Analysis}

We begin by considering the case where couplings decay with exponent
$\delta=2$. Figure~\ref{fig:Cmu_Cq_v_T_for_many_N} displays $\Cmu(N,T)$ and
$C_q(N,T)$---the $\Cmu$ and $C_q$ of processes $\SpinProcess(N,T)$---versus $T$
for $N=1, \ldots, 6$. The most striking feature is that the classical and
quantum memory requirements exhibit qualitatively very different behaviors.

Classical memory increases with $T$, saturating at $\Cmu = N$, since all
transitions become equally likely at high temperature. As a result there are
$2^N$ equally probable causal states and this means one needs $N$ bits of memory
to store the system's current state. For example, in the nearest-neighbor Ising
model (process $\SpinProcess(1,T)$) high temperature makes spin-$\uparrow$
and spin-$\downarrow$, and thus the corresponding states, equally
likely.\footnote{At $T = \infty$ these processes have only a single causal
state and thus $\Cmu=0$. This is a well known discontinuity that derives from
the sudden predictive-equivalence of all of the causal states there.}

Also, in the low-temperature limit, this system is known to yield one of only
two equally likely configurations---all spin-$\uparrow$ or all
spin-$\downarrow$. In other words, at low temperature
$p^{{\color{red}\pmb\downarrow}}_{\pmb{\uparrow}}$ and
$p^{{\color{red}\pmb\uparrow}}_{\pmb{\downarrow}}$ converge to zero, while
$p^{{\color{red}\pmb\uparrow}}_{\pmb{\uparrow}}$ and
$p^{{\color{red}\pmb\downarrow}}_{\pmb{\downarrow}}$ converge to
one.\footnote{It should be pointed out that at any finite temperature
$p^{{\color{red}\pmb\downarrow}}_{\pmb{\uparrow}}$ and
$p^{{\color{red}\pmb\uparrow}}_{\pmb{\downarrow}}$ are nonzero and, therefore,
the \eM\ states remain strongly-connected.} This is reflected in the
convergence of all curves at $\Cmu = 1$. Equivalently, this means one needs
only one bit of memory to store the current state.

We can similarly understand the qualitative behavior of $C_q(N,T)$ for a fixed
$N$. As temperature increases, all length-$N$ signal states become equivalent.
This is the same as saying that all length-$N$ spin configurations become
equally likely. As a consequence, the signal states approach one another and,
thus, $C_q(N,T)$ converges to zero.

In the low-temperature limit, the two $N$-$\uparrow$ and $N$-$\downarrow$
configurations are distinguished by the high likelihood of neighboring spins
being of like type. This leads to a von Neumann entropy of $S(\rho) = 1$.

\begin{figure}
\includegraphics[width=\linewidth]{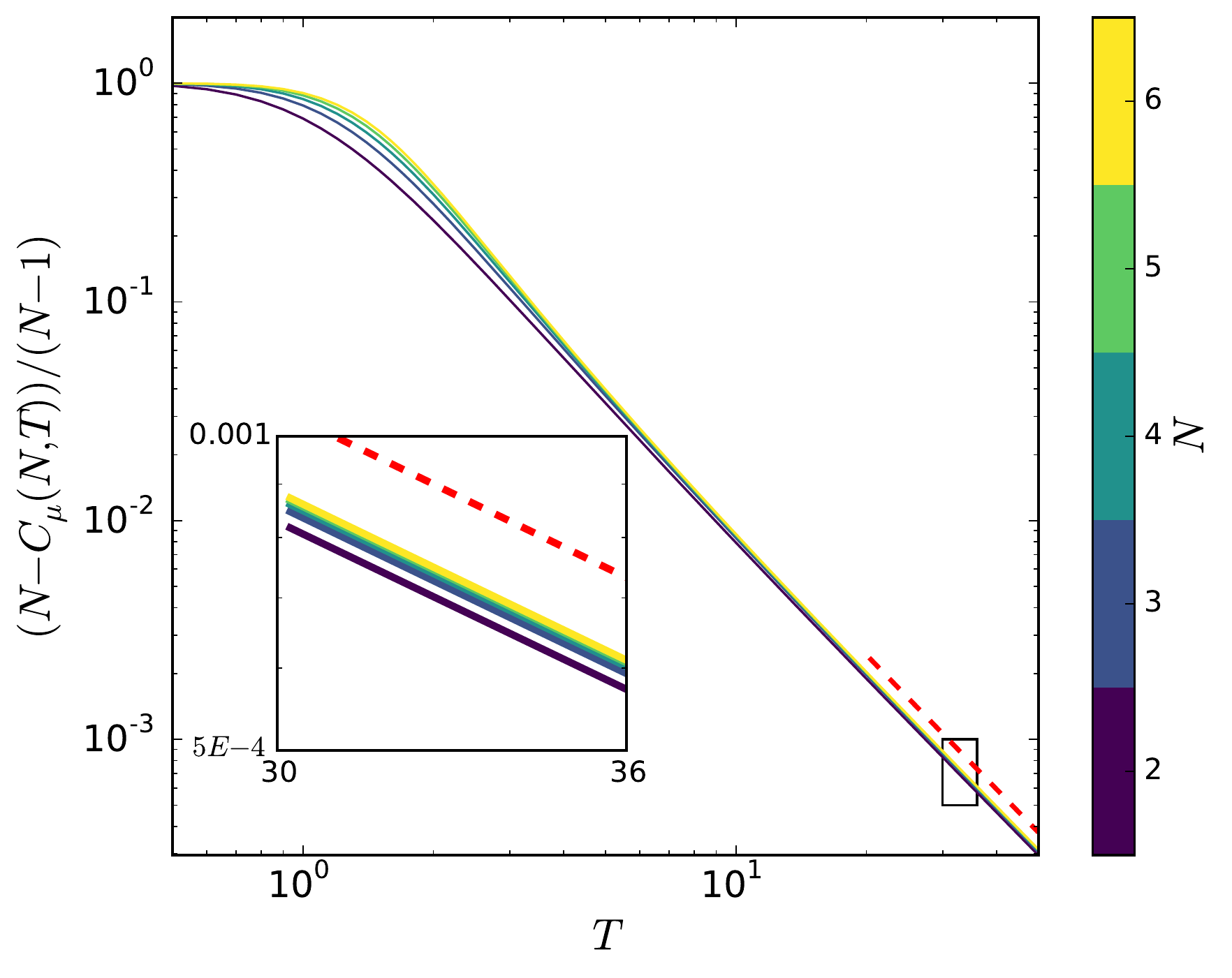}
\caption{Rescaling the classical memory requirement $\Cmu(N,T)$ to
	$(N-\Cmu)/(N-1)$ shows a tight data collapse, which is especially strong at
	high temperatures ($T > 2$). The asymptotic behavior is a power-law with
	scaling exponent $\gamma =2$. The inset zooms in to show $\Cmu$'s
	convergence with increasing $N$. While the figure shows the case
	$\delta=2$, but the slope $\gamma$ at high $T$ is independent of $T$.
	}
\label{fig:Cmu_v_T_for_many_N}
\end{figure}

Figure~\ref{fig:Cmu_Cq_v_T_for_many_N} reveals strong similarities in the form
of $\Cmu(T)$ at different $N$. A simple linear scaling leads to a substantial
data collapse, shown in Fig.~\ref{fig:Cmu_v_T_for_many_N}. The scaled curves
$(N-\Cmu)/(N-1)$ exhibit power-law behavior in $T$ for $T > 2$.
Increasing the temperature to $T=300$ (beyond the scale of the Fig.~\ref{fig:Cmu_v_T_for_many_N}) shows that this scaling is $\gamma \simeq 2.000$.
The scaling indicates how the classical memory saturates at high temperature.

This behavior is generic for different coupling decay values $\delta>1$ and,
more to the point, the scaling ($\gamma$) is independent of $\delta$. We do not
consider $\delta < 1$, where the system energy becomes nonextensive.

\begin{figure}
\includegraphics[width=\linewidth]{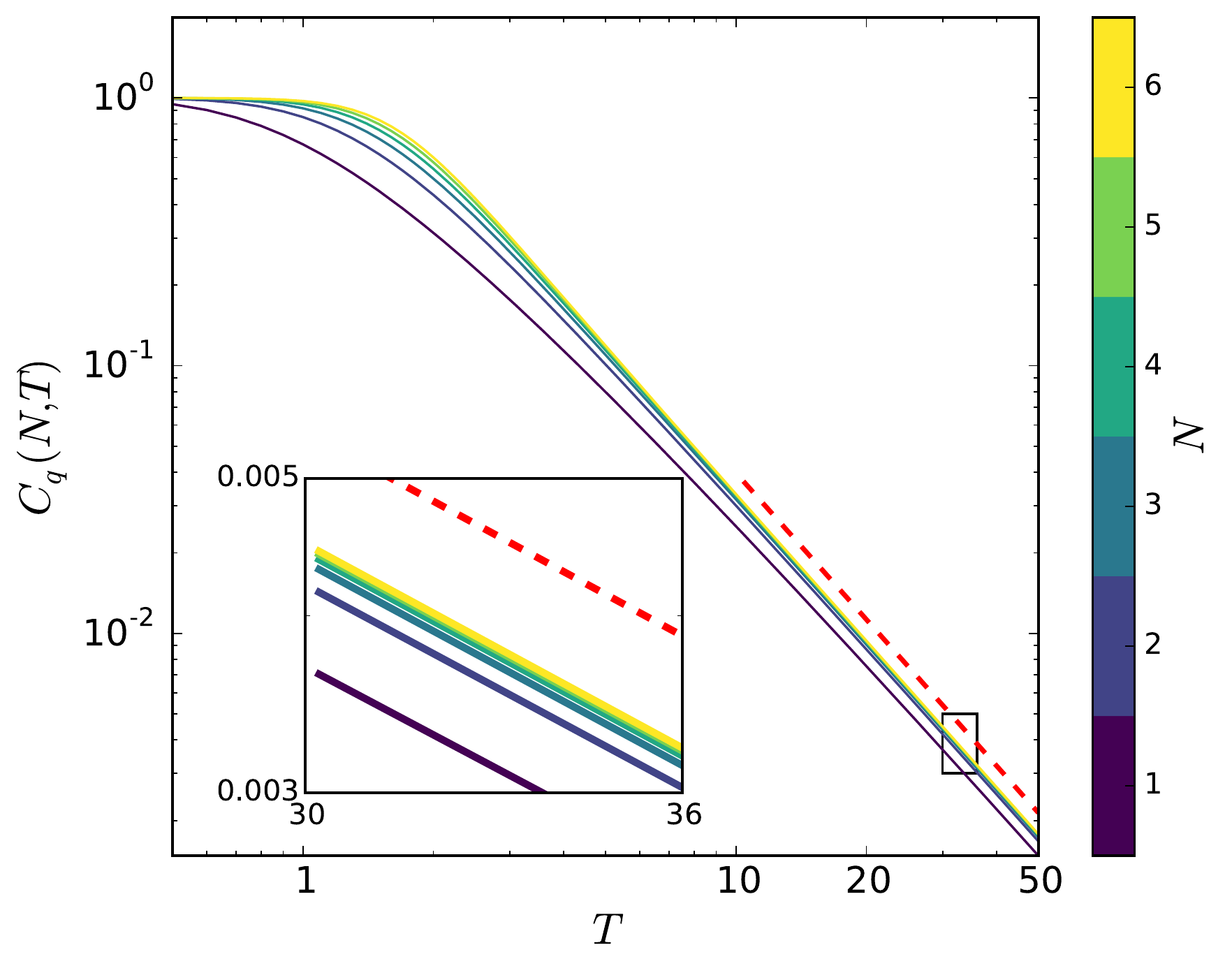}
\caption{Quantum memory $C_q(N,T)$ too shows a data collapse in $N$ that is
	also especially tight at high temperature ($T > 2$). The asymptotic
	behavior is a power-law with numerically estimated scaling exponent $\alpha
	=2$. The lower inset zooms to highlight convergence with increasing $N$.
	While the figure shows the case with $\delta=2$, but the slope $\alpha$ at
	high $T$ is independent of $T$.
	}
\label{fig:Cq_v_T_for_many_N}
\end{figure}

Now, we can analyze the decrease in $C_q$ with temperature.
Figure~\ref{fig:Cq_v_T_for_many_N} shows that $C_q$ is also a power-law in $T$.
By measuring this scaling exponent in the same way as above, we determine
$\alpha=2.000$.  Furthermore, analysis shows (App.~\ref{sec:HTB}) that:
\begin{align*}
C_q(N,T) \propto \frac{\log_2(T)}{T^{2}}
  ~.
\end{align*}
This verifies and adds detail to our numerical estimate.

This behavior is generic for different coupling decay values $\delta>1$ and,
moreover, the scaling exponent $\alpha$ is independent of $\delta$. It is
interesting to note that in this case no data-collapse rescaling is required.
The exponent $\alpha$ directly captures the extreme compactness of high
temperature quantum simulations.

Taking these results together, we can now appreciate the substantial relative
advantage of quantum versus classical simulators.

Define the \emph{quantum advantage} $\eta$ as the ratio of the minimum required
memory for the classical simulation to that for the quantum simulation:
\begin{align*}
\eta(N,T) \equiv \Cmu(N,T) / C_q(N,T)
  ~. 
\end{align*}
For fixed temperature $T\gtrapprox2$, $\Cmu(N,T)$ is approximately linear in
$N$ and for a fixed $N$ it is approximately independent of $T$.
As a consequence, the asymptotic quantum advantage scales as:
\begin{align*}
\eta(N,T) \propto N \frac{T^{2}}{\log_2(T)}
  ~,
\end{align*}
which increases faster than any $T^r$ for $r<2$. Thus, the answer to our
motivating question is that the quantum advantage does in fact display a
scaling behavior, increasing without bound with interaction range $N$ and even
faster with temperature $T$.

\section{Conclusion}

It is notoriously hard to find quantum advantage and even harder to prove
\cite{Dale15a}. Here, we found just such an advantage in the realm of
stochastic-process simulation. Concretely, we analyzed the $N$-nearest neighbor
Ising spin system and demonstrated that its quantum advantage displays generic
scaling behavior---quadratic in temperature and linear in interaction range.
What does this mean? The most striking conclusion is that a highly interacting
classical system can be simulated with unbounded quantum advantage. Given the
simplicity of the Ising system, we conjecture that this scaling behavior may be
a universal feature of quantum advantage in the simulation of extended physical
systems.

The Ising model has contributed great insights to condensed matter physics,
however, it is classical. Given our examining the difference between classical
and quantum simulators, it is natural to wonder about this difference in the
context of a truly quantum Hamiltonian. Is the quantum advantage amplified?
Are there systems for which we find no quantum advantage? What is the
distinguishing characteristic?

For finite-range interaction in one dimension, there is no Ising phase
transition. How might the quantum advantage change in the presence of such a
transition? Given that the quantum advantage here scaled with the interaction
range, we might expect that near the critical temperature, where long-range
interactions are important, the quantum advantage is amplified.

\section*{Acknowledgments}
\label{sec:acknowledgments}

We thank Mehrnaz Anvari for her suggestions and the Santa Fe Institute for its
hospitality during visits. This material is based upon work supported by, or in
part by, the John Templeton Foundation and the U. S. Army Research Laboratory
and the U. S. Army Research Office under contracts W911NF-13-1-0390 and
W911NF-13-1-0340.

\bibliography{chaos,ref}

\begin{thebibliography}{10}

\bibitem{Whit78a}
A.~N. Whitehead.
\newblock {\em Process and Reality}.
\newblock The Free Press, New York, corrected edition, 1978.

\bibitem{Ross06a}
D.~Ross, J.~Ladyman, D.~Spurrett, and J.~Collier.
\newblock {\em Everything Must Go: Metaphysics Naturalized}.
\newblock Oxford University Press, 2007.

\bibitem{Ande89a}
P.~W. Anderson.
\newblock Spin glass {VI}: {Spin} glass as cornucopia.
\newblock {\em Physics Today}, September:9--11, 1989.

\bibitem{Stei13a}
D.~K. Stein and C.~M. Newman.
\newblock {\em Spin Glasses and Complexity}.
\newblock Princeton University Press, Princeton, New Jesey, 2013.

\bibitem{Feyn82a}
R.~Feynman.
\newblock Simulating physics with computers.
\newblock {\em Intl. J. Theo. Phys.}, 21(6/7):467--488, 1982.

\bibitem{Maho16}
J.~R. Mahoney, C.~Aghamohammadi, and J.~P. Crutchfield.
\newblock Occam's quantum strop: Synchronizing and compressing classical
  cryptic processes via a quantum channel.
\newblock {\em Scientific Reports}, 6, 2016.

\bibitem{BAXTER07}
R.~J. Baxter.
\newblock {\em Exactly solved models in statistical mechanics}.
\newblock Courier Corporation, 2007.

\bibitem{Agha12}
A.. Aghamohammadi, C.~Aghamohammadi, and M.~Khorrami.
\newblock Externally driven one-dimensional ising model.
\newblock {\em J. Stat. Mechanics: Theory and Experiment}, 2012(02):P02004,
  2012.

\bibitem{Dyson69}
F.~J. Dyson.
\newblock Existence of a phase-transition in a one-dimensional {Ising}
  ferromagnet.
\newblock {\em Comm. Math. Physics}, 12(2):91--107, 1969.

\bibitem{Uppe97a}
D.~R. Upper.
\newblock {\em Theory and Algorithms for Hidden {M}arkov Models and Generalized
  Hidden {M}arkov Models}.
\newblock PhD thesis, University of California, Berkeley, 1997.
\newblock {P}ublished by University Microfilms Intl, Ann Arbor, Michigan.

\bibitem{Travers13}
N.~F. Travers.
\newblock {\em Bounds on Convergence of Entropy Rate Approximations in Hidden
  {Markov} Processes}.
\newblock PhD thesis, University of California, Davis, 2013.

\bibitem{Crut97a}
J.~P. Crutchfield and D.~P. Feldman.
\newblock Statistical complexity of simple one-dimensional spin systems.
\newblock {\em Phys. Rev. E}, 55(2):R1239--R1243, 1997.

\bibitem{Feld98b}
D.~P. Feldman and J.~P. Crutchfield.
\newblock Discovering non-critical organization: {S}tatistical mechanical,
  information theoretic, and computational views of patterns in simple
  one-dimensional spin systems.
\newblock Santa Fe Institute, 1998.
\newblock {Santa} Fe Institute Working Paper 98-04-026.

\bibitem{Norr98}
J.~R. Norris.
\newblock {\em Markov Chains}, volume~2.
\newblock Cambridge University Press, 1998.

\bibitem{LEVIN09}
D.~A. Levin, Y.~Peres, and E.~L. Wilmer.
\newblock {\em Markov Chains and Mixing Times}.
\newblock Am. Math. Soc., 2009.

\bibitem{Rabi86a}
L.~R. Rabiner and B.~H. Juang.
\newblock An introduction to hidden {Markov} models.
\newblock {\em IEEE ASSP Magazine}, January, 1986.

\bibitem{Rabi89a}
L.~R. Rabiner.
\newblock A tutorial on hidden {Markov} models and selected applications.
\newblock {\em IEEE Proc.}, 77:257, 1989.

\bibitem{Marz14b}
S.~Marzen and J.~P. Crutchfield.
\newblock Informational and causal architecture of discrete-time renewal
  processes.
\newblock {\em Entropy}, 17(7):4891--4917, 2015.

\bibitem{Lohr09b}
W.~L{\"o}hr and N.~Ay.
\newblock Non-sufficient memories that are sufficient for prediction.
\newblock In {\em International Conference on Complex Sciences}, pages
  265--276. Springer, 2009.

\bibitem{Lohr09c}
W.~L{\"o}hr and N.~Ay.
\newblock On the generative nature of prediction.
\newblock {\em Adv. Complex Sys.}, 12(02):169--194, 2009.

\bibitem{Lohr12}
W.~L{\"o}hr.
\newblock Predictive models and generative complexity.
\newblock {\em J. Systems Sci. Complexity}, 25(1):30--45, 2012.

\bibitem{Gmei11a}
P.~Gmeiner.
\newblock Equality conditions for internal entropies of certain classical and
  quantum models.
\newblock {\em arXiv preprint arXiv:1108.5303}, 2011.

\bibitem{Crut12a}
J.~P. Crutchfield.
\newblock Between order and chaos.
\newblock {\em Nature Physics}, 8(1):17--24, 2012.

\bibitem{Crut88a}
J.~P. Crutchfield and K.~Young.
\newblock Inferring statistical complexity.
\newblock {\em Phys. Rev. Let.}, 63:105--108, 1989.

\bibitem{Perr99a}
N.~Perry and P.-M. Binder.
\newblock Finite statistical complexity for sofic systems.
\newblock {\em Phys. Rev. E}, 60:459--463, 1999.

\bibitem{Delg97a}
J.~Delgado and R.~V. Sol{\'e}.
\newblock Collective-induced computation.
\newblock {\em Phys. Rev. E}, 55:2338--2344, 1997.

\bibitem{Neru10}
D.~Nerukh, C.~H. Jensen, and R.~C. Glen.
\newblock Identifying and correcting {non-Markov} states in peptide
  conformational dynamics.
\newblock {\em J. Chem. Physics}, 132(8):084104, 2010.

\bibitem{Nerukh12}
D.~Nerukh.
\newblock {Non-Markov} state model of peptide dynamics.
\newblock {\em J. Mole. Liquids}, 176:65--70, 2012.

\bibitem{Kell12a}
D.~Kelly, M.~Dillingham, A.~Hudson, and K.~Wiesner.
\newblock A new method for inferring hidden {M}arkov models from noisy time
  sequences.
\newblock {\em PLoS One}, 7(1):e29703, 01 2012.

\bibitem{Li13a}
C.-B. Li and T.~Komatsuzaki.
\newblock Aggregated {M}arkov model using time series of a single molecule
  dwell times with a minimum of excessive information.
\newblock {\em Phys. Rev. Lett.}, 111:058301, 2013.

\bibitem{Varn14a}
D.~P. Varn and J.~P. Crutchfield.
\newblock Chaotic crystallography: {How} the physics of information reveals
  structural order in materials.
\newblock {\em Curr. Opin. Chem. Eng.}, 7:47--56, 2015.

\bibitem{Crut13a}
J.~P. Crutchfield, P.~Riechers, and C.~J. Ellison.
\newblock Exact complexity: {Spectral} decomposition of intrinsic computation.
\newblock {\em Phys. Lett. A}, 380(9-10):998--1002, 2016.

\bibitem{Maho09a}
J.~R. Mahoney, C.~J. Ellison, and J.~P. Crutchfield.
\newblock Information accessibility and cryptic processes.
\newblock {\em J. Phys. A: Math. Theo.}, 42:362002, 2009.

\bibitem{Maho11a}
J.~R. Mahoney, C.~J. Ellison, R.~G. James, and J.~P. Crutchfield.
\newblock How hidden are hidden processes? {A} primer on crypticity and entropy
  convergence.
\newblock {\em CHAOS}, 21(3):037112, 2011.

\bibitem{Cove06a}
T.~M. Cover and J.~A. Thomas.
\newblock {\em Elements of Information Theory}.
\newblock Wiley-Interscience, New York, second edition, 2006.

\bibitem{Dobs69}
J.~F. Dobson.
\newblock Many-neighbored {Ising} chain.
\newblock {\em J. Math. Physics}, 10(1):40--45 

\bibitem{Gu12a}
M.~Gu, K.~Wiesner, E.~Rieper, and V.~Vedral.
\newblock Quantum mechanics can reduce the complexity of classical models.
\newblock {\em Nature Comm.}, 3:762, 2012.

\bibitem{Riech16a}
P.~M. Riechers, J.~R. Mahoney, C.~Aghamohammadi, and J.~P. Crutchfield.
\newblock Minimized state complexity of quantum-encoded cryptic processes.
\newblock {\em Phys. Rev. A}, 93(5):052317, 2016.

\bibitem{Monr16}
A.~Monras and A.~Winter.
\newblock Quantum learning of classical stochastic processes: The completely
  positive realization problem.
\newblock {\em J. Math. Physics}, 57(1):015219 

\bibitem{Monr10a}
A.~Monras, A.~Beige, and K.~Wiesner.
\newblock Hidden quantum {Markov} models and non-adaptive read-out of many-body
  states.
\newblock {\em arXiv:1002.2337}, 2010.

\bibitem{Agha16a}
C.~Aghamohammadi, J.~R. Mahoney, and J.~P. Crutchfield.
\newblock The ambiguity of simplicity.
\newblock {\em arXiv:1602.08646}, 2016.

\bibitem{Suen15a}
W.~Y. Suen, J.~Thompson, A.~J.~P. Garner, V.~Vedral, and M.~Gu.
\newblock The classical-quantum divergence of complexity in the {Ising} spin
  chain.
\newblock {\em arXiv preprint arXiv:1511.05738}, 2015.

\bibitem{Tan14}
R.~Tan, D.~R. Terno, J.~Thompson, V.~Vedral, and M.~Gu.
\newblock Towards quantifying complexity with quantum mechanics.
\newblock {\em Euro. Phys. J. Plus}, 129(9):1--12, 2014.

\bibitem{Pals15}
M.~S. Palsson, M.~Gu, J.~Ho, H.~M. Wiseman, and G.~J. Pryde.
\newblock Experimental quantum processing enhancement in modelling stochastic
  processes.
\newblock {\em arXiv:1602.05683}, 2015.

\bibitem{Dale15a}
H.~Dale, D.~Jennings, and T.~Rudolph.
\newblock Provable quantum advantage in randomness processing.
\newblock {\em Nature Comm.}, 6, 2015.

\bibitem{Rush48}
G.~S. Rushbrooke and H.~D. Ursell.
\newblock On one-dimensional regular assemblies.
\newblock In {\em Math. Proc. Cambridge Phil. Soc.}, volume~44, pages 263--271.
  Cambridge University Press, 1948.

\bibitem{Froh82}
J.~Fr{\"o}hlich and T.~Spencer.
\newblock The phase transition in the one-dimensional {Ising} model with 1/r 2
  interaction energy.
\newblock {\em Comm. Math. Physics}, 84(1):87--101, 1982.

\bibitem{Fisher72}
M.~E. Fisher, S.~Ma, and B.~G. Nickel.
\newblock Critical exponents for long-range interactions.
\newblock {\em Phys. Rev. Let.}, 29(14):917, 1972.

\bibitem{Bianc13}
T.~Blanchard, M.~Picco, and M.~A. Rajabpour.
\newblock Influence of long-range interactions on the critical behavior of the
  {Ising} model.
\newblock {\em Europhysics Lett.}, 101(5):56003, 2013.

\end{thebibliography}

\cleardoublepage

\appendix

\section{Why the Dyson model?}
\label{sec:dyson_phase_transition}

The ferromagnetic Ising spin linear chain with finite-range interaction cannot undergo a phase transition at any positive temperature \cite{Rush48}. In contrast, the Dyson model has a standard second-order phase transition for a range of $\delta$. It was analytically proven by Dyson \cite{Dyson69} that a phase transition exists for $1 < \delta < 2$. The existence of a transition at $\delta=2$ was proven much later on \cite{Froh82}. It is also known that there exists no phase transition for $\delta>3$ \cite{Fisher72} where it behaves as a short-range system. Finally, it was demonstrated numerically that the parameter regime $2<\delta \leq 3$ contains a phase transition \cite{Bianc13}, however, this fact has resisted analytical proof. For $\delta \leq 1$, the model is considered nonphysical since the energy becomes non-extensive.

For these reasons we selected the Dyson spin model as it provides the
simplicity of 1D configurations, while generating nontrivially correlation spin
configurations.

\section{\EM Construction}
\label{sec:AnalyticalResult}

We show how to construct the \eM\ simulator of the process $\SpinProcess(N,T)$.
Consider a block of spins of length $2N$, divided equally into two blocks.  We
denote spins in the left (L) and right (R) halves by: $s_{i}^{L}$ and
$s_{i}^{R}$ for $i=1, \cdots N$, respectively. We map the left and right block
configurations each to an integer $\eta_X$ by:
\begin{align*}
\eta_{\ast}= \sum_{i=1}^{N}
  \left( \frac{s_{i}^{\ast}+1}{2} \right) 2^{i-1}
  ~,
\end{align*}
$\ast \in \{L, R\}$. The blocks internal energies are given by:
\begin{align*}
X_{\eta_{\ast}}= -B \sum_{i=1}^{N} s_{i}^{\ast}  -\sum_{i=1}^{N-1} \sum_{k=1}^{N-i} J_i s_{k}^{\ast} s_{k+i}^{\ast}~,
\end{align*}
and the correlated energy between two blocks is:
\begin{align*}
Y_{\eta_L,\eta_R}= -\sum_{i=1}^{N} \sum_{k=1}^{i} J_i s_{N-k+1}^{L} s_{k}^{R}~.
\end{align*}
With these we construct the transfer matrix:
\begin{align*}
V_{\eta_L,\eta_R} =
  \exp{
  \left(
  -\frac{1}{T}( \half X_{\eta_L} + Y_{\eta_L,\eta_R} + \half X_{\eta_R} )
  \right)
  }
  ~.
\end{align*}
Reference \cite{Feld98b} shows that the \eM\ labeled-transition matrices can be
written as:
\begin{align*}
T^{(x)}_{\eta_0,\eta_1} = 
\begin{cases}
\frac{1}{\lambda} V_{\eta_0,\eta_1} \frac{u_{\eta_1}}{u_{\eta_0}} \quad &\eta_1=(\lfloor \frac{\eta_0}{2} \rfloor + x*2^{N-1}) \\
0  \quad &\eta_1\neq(\lfloor \frac{\eta_0}{2} \rfloor + x*2^{N-1})
\end{cases}
  ~.
\end{align*}
Then the \eM\ simulator of $\SpinProcess(N,T)$ is $\big\{ \CausalStateSet ,
\,\MeasAlphabet, \, \{ T^{(\meassymbol)} \}_{ \meassymbol \in \MeasAlphabet}
\big\}$ where $\MeasAlphabet=\{ 0,1 \}$ and $\CausalStateSet=\{i: 1\leq i \leq
2^N\}$.

\begin{figure}
\includegraphics[width=\linewidth]{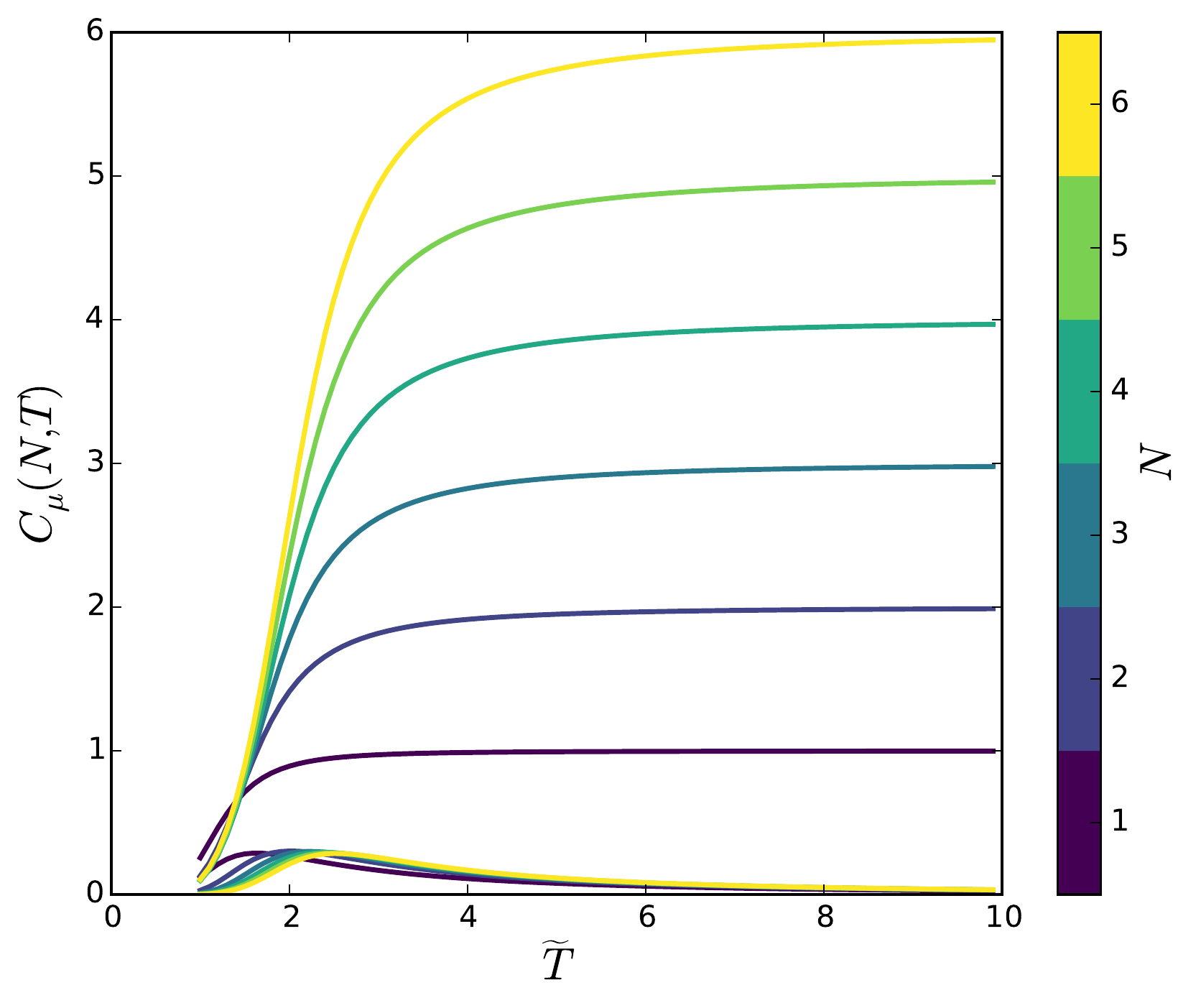}
\caption{Magnetic field effects on classical $\Cmu(N,T)$ and quantum memory
	$C_q(N,T)$ requirements for simulating the process generated by Hamiltonian
	$\mathcal{\widehat{H}}_N$ for $N=1, \ldots, 6$ and a range of temperatures
	$T = 1, \ldots, 50$ at $B=0.3$.
  }
\label{fig:Cmu_Cq_v_T_for_many_N_B}
\end{figure}

\section{Presence of Magnetic Field }
\label{sec:PMF}

Naturally, one might ask how our results are modified by the presence of an
external magnetic field. Consider the one-dimensional ferromagnetic Ising spin
chain with Hamiltonian:
\begin{align*}
\mathcal{\widehat{H}}_N =
  -\sum_{i} \sum_{k=1}^{N}\frac{J_0}{{k}^\delta} s_is_{i+k} - \sum_{i} Bs_i
  ~.
\end{align*}
 
Figure~\ref{fig:Cmu_Cq_v_T_for_many_N_B} shows that, due to the symmetry
breaking at low temperature, both $C_q(N,T)$ and $\Cmu(N,T)$ converge to zero.
(All the spins at low temperature align with magnetic field and, as a
consequence, no memory is needed.) The high temperature behaviors for both
functions are the same as before and the quantum advantage remains the same.

\section{High Temperature Behavior}
\label{sec:HTB}

Consider first the case $N=1$. Due to symmetry we have $p \equiv
\Pr(\uparrow | \uparrow) = \Pr(\downarrow | \downarrow) = N/D$,
where $N = \exp{(\beta J)}$ and $D = \exp{(\beta J)}+
\sqrt{\exp{(-2\beta J)}}$ with $\beta = 1/T$.  At high temperature
$\beta$ is small and we have:
\begin{align*}
D &= 2+ \beta^2,\\
N &= 1+ \beta + \beta ^2 
~.
\end{align*}

Again, due to symmetry we have $\pi_1=\pi_2=\half$. Therefore, the
density matrix in Eq.~(\ref{rho}) is:
\begin{align*}
\rho=\left(\begin{matrix}
\half & \sqrt{p(1-p)}\\ & \\\sqrt{p(1-p)}&\half
 \end{matrix}\right)\nonumber
~,
\end{align*}
which has two eigenvalues $\beta^2 / 4$ and $1 - \beta^2 / 4$.
As a consequence, $C_q$---$\rho$'s von Neumann entropy---is:
\begin{align*}
C_q & = S(\rho) \\
    & \simeq -\left( \frac{\beta^2}{4} \log_2 \frac{\beta^2}{4}
	+ \left(1-\frac{\beta^2}{4}\right)
	\log_2 \left(1-\frac{\beta^2}{4}\right) \right) \\
    & \simeq \frac{\log_2{(T)}}{2T^2}
  ~.
\end{align*}
Examining the numerator, for any $r>0$ we have $\log_2{(T)}<T^r$.
So, for large $T$ and for all $r > 0$:
\begin{align*}
\frac{1}{T^2} < \frac{\log_2{(T)}}{T^2} < \frac{1}{T^{r+2}}
~.
\end{align*}
This explains the fat tails of $C_q$ for large $T$. More to the
point, it shows that for $N=1$ the scaling exponent is $\alpha = 2$. 

Increasing the temperature, the interaction between spins weakens.
At high temperature the only important neighbor is the nearest neighbor.
And so, the high-temperature behavior is similar to the case of $N=1$ and is independent of $N$.

\end{document}